\documentclass[useAMS,usenatbib]{mn2e}

\usepackage{longtable}
\usepackage{pdflscape}
\usepackage{graphicx}
\usepackage{pdflscape}
\usepackage{lscape,graphicx,pifont}
\usepackage{color}

\newcommand{\Mdot}[2]{\mbox{${#1}\times10^{-{#2}}$\,M$_\odot$~yr$^{-1}$}}
\newcommand{\Msun}{\mbox{\,M$_\odot$}}
\newcommand{\Rsun}{\mbox{\,R$_\odot$}}
\newcommand{\Lsun}{\mbox{\,L$_\odot$}}
\newcommand{\vunit}{\mbox{\,km\,s$^{-1}$}}
\newcommand{\mic}{\mbox{$\,\mu$m}}
\newcommand{\pion}[2]{{#1}\,{\sc {#2}}}
\newcommand{\fion}[2]{[{#1}\,{\sc {#2}}]}

\newcommand{\ltsimeq}{\raisebox{-0.6ex}{$\,\stackrel
        {\raisebox{-.2ex}{$\textstyle <$}}{\sim}\,$}}

\newcommand{\us}{\mbox{U~Sco}}

\newcommand{\ncrit}{\mbox{$n_{\rm crit}$}}

\usepackage{graphicx}	
\usepackage{amsmath}	
\usepackage{amssymb}	

\title[IR spectroscopy of U~Sco]{Infrared spectroscopy of the 
2022 eruption of the recurrent nova U~Sco}

\author[A. Evans et al]{A. Evans$^1$\thanks{E-mail:a.evans@keele.ac.uk},
D. P. K. Banerjee$^2$, 
C. E. Woodward$^3$\thanks{Visiting Astronomer at the Infrared Telescope 
Facility, which is operated by the University of Hawaii under contract
80HQTR19D0030 with the National Aeronautics and Space Administration.},
T. R. Geballe$^4$,
R. D. Gehrz$^3$,  \newauthor 
K. L. Page$^5$,
S. Starrfield$^6$
 \\ 
$^1$Astrophysics Group, Lennard Jones Laboratory, Keele University, Keele, Staffordshire,  ST5 5BG, UK\\ 
$^2$Physical Research Laboratory, Navrangpura,  Ahmedabad, Gujarat 
380009, India\\ 
$^3$Minnesota Institute for Astrophysics, School of Physics \& Astronomy,
116 Church Street SE, University of Minnesota, \\
Minneapolis, MN 55455, USA\\ 
$^4$Gemini Observatory/NSF's NOIRLab, 670 N. Aohoku Place, Hilo, HI, 96720,
USA\\ 
$^5$School of Physics and Astronomy, University of Leicester, University Road, Leicester, LE1 7RH, UK \\ 
$^6$School of Earth and Space Exploration, Arizona State University, 
Box 871404, Tempe, AZ 85287-6004, USA\\ 
}

\date{Accepted XXX. Received YYY; in original form ZZZ}

\pubyear{2023}

\begin{document}
\label{firstpage}
\pagerange{\pageref{firstpage}--\pageref{lastpage}}
\maketitle

\begin{abstract}
We present near-infrared spectroscopy of the 2022 eruption
of the recurrent nova \us, over the period from 5.4 to 45.6~days after 
outburst. This is the {most intensive infrared 
study of this nova. Our observations started early after the outburst
and extended} almost to the end of the ``Super Soft'' X-ray phase. 
A major find is the presence of coronal lines from day~9.41, 
{one of
the earliest appearances of these in any nova, classical or recurrent.}
The temperature of 
the coronal gas is $7\times10^5$~K. There is also evidence for 
the presence of much cooler ($\ltsimeq2.5\times10^4$~K) gas. 
Remarkable changes are seen in the \pion{He}{i} 1.083\mic\ line,
the strength of which declines, then recovers, in anti-correlation
with the X-ray behaviour. 
We conclude that shock ionisation is the dominant excitation 
mechanism for the coronal line emission.
There is evidence in the infrared spectra for the presence of  
black body emission at $\sim20000$~K, 
{which we tentatively identify with the irradiated secondary,} 
and for free-free/free-bound emission.
For the previously determined binary inclination of $82\fdg7$, the 
implied ejection velocities are as high as 22000\vunit.
These velocities appear unprecedented in nova outflows, and are
comparable to those seen in supernovae, thereby marking \us\ 
as a truly remarkable object.
\end{abstract}

\begin{keywords}
stars: individual: U~Sco ---
novae, cataclysmic variables ---
infrared: stars 
\end{keywords}



\section{Introduction}
Recurrent novae (RNe) are a subset of the cataclysmic variable
binary systems in which a cool component (the secondary)
transfers matter via an accretion disk (AD) onto a white dwarf 
(WD; the primary). The base of the layer accreted on the WD becomes 
degenerate, and eventually hot enough to initiate a thermonuclear runaway, 
resulting in a nova eruption. In time, accretion resumes and eventually 
another eruption occurs. All novae are recurrent but the eruptions of 
RNe repeat on timescales $\ltsimeq100$~yr 
\citep[see][for a review]{anupama08}.

\section{\us: nature and recent eruptions}
\subsection{The binary}

\us\ is the most frequent eruptor of the known Galactic RNe. 
It is known to have undergone eruptions in 1863, (1873, 1884, 1894), 
1906, 1917, (1927), 1936, {1945}, (1955), 
{1969}, 1979, 1987, 1999, 2010 (2016)
\citep[dates in brackets are probable eruptions; see][]{darnley21,schaefer22}.
\us\ is an eclipsing binary, consisting of a $1.55\pm0.24$\Msun\ WD 
and a $0.88\pm0.17$\Msun\ secondary; its inclination is $i = 82\fdg7\pm2\fdg9$ \citep{thoroughgood01}.
The orbital period prior to the 2022 eruption was
$1.2305658\pm0.0000041$~days, as measured from 2016.78 
(the time of the putative 2016  eruption) to 2022.4
\citep{schaefer22}. \citeauthor{schaefer22}
also found an increase of 22.4~parts-per-million in 
the orbital period across the 2010 eruption,
concluding that the period change is most
likely due to the reaction of the binary to the asymmetric ejection 
of material during the nova eruption.

While the uncertainty is relatively large, the mass of the WD in \us\
seems uncomfortably high, although the mass limit for
differentially rotating white dwarfs can be as high as 4\Msun, 
i.e., far higher than the 
canonical Chandrasekhar limit \citep[e.g.,][]{yoon05}.
\cite{hachisu00} modelled the light curve from the 1999 eruption, and
found a WD mass of $1.37\pm0.01$\Msun, very close to that found by
\cite{shara18} on the basis of modelling the eruptions of classical
and RNe. We assume 1.37\Msun\ here.
The secondary star is a sub-giant of spectral type K2
\citep{anupama08}. Therefore, unlike RN systems such as RS~Oph, 
in which the secondary is a red giant with a strong wind, the 
secondary in \us\ is evolved but has little or no wind. 
For this reason the evolution of an
eruption in \us\ is expected to differ significantly from that in a 
RN system with a giant secondary.

There is evidence that less mass is ejected during an eruption 
of \us\ than is 
accreted by the WD between outbursts \citep{kahabka99}, 
suggesting that the WD mass in \us\ is increasing. 
{As the WD likely has CO composition
\citep{mason13}, \us\ is a candidate Type~Ia supernova progenitor
\citep{starrfield20}.}

The distance of \us\ is poorly constrained \citep{schaefer10}, but is known to
be large, between $\sim8.5$~kpc and $\sim15.4$~kpc. Where we need to
use a distance here, we use 10~kpc.

\label{2022}
\begin{figure}
 \includegraphics[width=9cm,keepaspectratio]{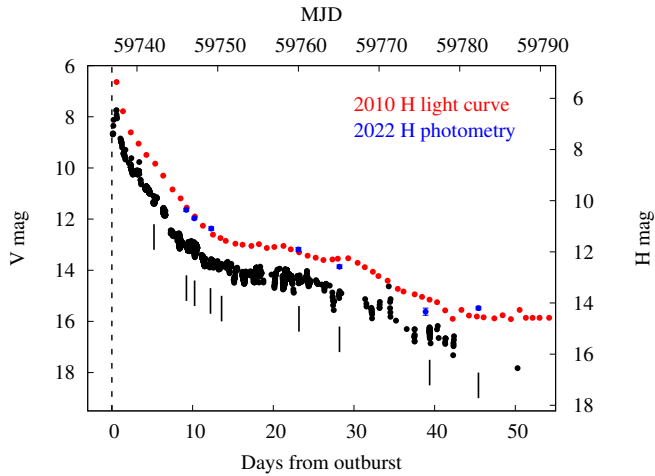}
\caption{Black points: $V$ band light curve of 2022 eruption; 
AAVSO data$^1$.
The dashed vertical line indicates the assumed time origin, MJD 59736.72 (see text).
Times of the observations described here are indicated by short vertical lines.
Red points: $H$ band light curve of the 2010 eruption \citep[][]{pagnota15}. 
Blue points: $H$ photometry as described in text for the 2022
eruption.\label{LC}}
\end{figure}

\subsection{A brief summary of three recent eruptions}

The 1979 eruption of \us\ was observed at ultraviolet wavelengths by
\cite{williams81}, and at optical and ultraviolet wavelengths
by \cite{barlow81}. While both studies noted the presence of high 
ionisation states, such as \pion{C}{iv}, \pion{N}{v}, \pion{O}{iv}, 
\pion{O}{vi}, they did not report any forbidden lines.
Indeed, \cite{anupama08} stated that, up until the 1999 eruption, no 
forbidden lines had been reported during outburst.
\citeauthor{barlow81} and \citeauthor{williams81} 
also estimated that the ejecta mass was low, $\sim10^{-7}$\Msun,
and that there was evidence for an overabundance of He relative to H.

The 1999 eruption was observed in the optical by \cite{munari99}.
These authors reported a narrowing of H$\,\alpha$, implying
a deceleration of $\sim270$\vunit~day$^{-1}$, and an increase in the 
degree of ionisation as the eruption progressed.
\cite{anupama13} found short-term spectral variations during the 2010
eruption, and reported intrinsic linear polarisation due to electron
scattering from the disc and jets. They estimated the mass of
ejected hydrogen was $\sim4.6\times10^{-6}$\Msun. 
\cite{maxwell12} determined that the helium abundance in the ejecta
was not significantly different from solar, 
{in contrast to previous 
estimates, which had $N(\mbox{He})/N(\mbox{H})$ ranging 
from 0.16 to 4.5 \citep{evans01,iijima02}.} In an erratum, \cite{maxwell15}
gave $N(\mbox{He})/N(\mbox{H})=0.117\pm0.014$.
All post-1979 eruptions have been characterised by 
high ejection velocities 
{\citep[$\sim10000$\vunit; e.g.,][]{barlow81,williams81,
iijima02,anupama13}.}

\cite{drake10} carried out 3-dimensional hydrodynamic simulations of the 
2010 eruption. They found that, as the ejecta encountered
the secondary star, the secondary gave rise to a ``bow shock'', 
in which the ejecta were shocked to a temperature of a few $\times10^5$~K.
They further  concluded that high wind and equatorial gas 
densities result in X-ray emission that exceeded the observational
upper limits, and that the circumstellar gas density is therefore rather
low. \cite{hachisu00} modelled the light curve of the 1999 eruption, and
concluded that the ``burning'' phase of the eruption, which is 
accompanied by a strong wind, persisted for the first $\sim20$~days
of the eruption. In this case one might expect the bow shock associated
with the secondary to persist for that length of time.

Near-infrared (NIR) observations were obtained during the eruptions
of 1999 \citep{evans01} and 2010 \citep{banerjee10,mason12,maxwell12,rudy22}.
\cite{evans01} noted the presence of an excess in the 
continuum, which they attributed to free-free emission.
\citeauthor{evans01} reported that there were no coronal lines in the 
1999 eruption spectrum, but it is likely that the strong line at 
$\sim2.05$\mic\ in their data was \fion{Al}{ix} $\lambda=2.045$\mic\ 
rather than \pion{He}{i} 2.059\mic\ (see below). 
{This is also likely to be the case for the 
line at approximately the same wavelength seen in the 2010 eruption
and tentatively identified as \pion{He}{i} by \cite{mason12}.}
Following the report 
of coronal line emission in the 2022 eruption \citep{banerjee22}, 
\cite{rudy22} reported re-examinations of their spectra of the 
2010 outburst, and those of other observers 
\citep[e.g.,][]{banerjee10, mason12, maxwell12}. 
They noted the presence of coronal lines in the 2010 outburst as well,
which either had not been identified, or had gone unnoticed.

\subsection{The 2022 eruption}

\begin{table*}
\caption{Observing log. \label{log}}
\begin{tabular}{cccccccccc}\hline
2022 Date &     MJD    &  Instrument & Resolution$^*$ & $\lambda$ range & \multicolumn{2}{c}{Airmass} &   Day$^\dag$  & $H$ & $T_{\rm BB}$ \\
          &            &            &  ($\lambda/\Delta\lambda$)   & (\mic)    & \us  & Standard&         &   (mag)          & (K)   \\\hline
June 12   &   59742.10 & F2   &  200-1200 & 0.9--2.5 & 1.06 & 1.08 &  5.38 &  ---            & ---  \\        
June 16   &   59746.13 & F2   &  200-1200 & 0.9--2.5 & 1.03 & 1.04&   9.41 & $11.64\pm0.06$  & $19000\pm2000$   \\ 
June 17   &   59747.14 & F2   &  200-1200 & 0.9--2.5 & 1.25&1.19&  10.42 & $11.97\pm0.05$  & $16500\pm1000$   \\
          &   59747.14 & F2   & 1700-3200 & 2.00--2.32  & 1.23&1.19  &  $''$     &         &                   \\
June 19   &   59749.11 & F2   & 200-1200  & 0.9--2.5 &1.03&1.03 &  12.39 & $12.37\pm0.07$  & ---  \\      
          &   59749.11 & F2   & 1700-3200 & 2.00--2.32 &1.03 &1.03&  $''$      &          &             \\    
June 20   &   59750.47 & SpeX &  1200      &  0.7--2.55  & 1.65 & 1.55 & 13.75 &  ---            & --- \\ 
June 30   &   59760.08 & F2   & 200-1200  & 0.9--2.5  &1.03&1.03 &   23.36 & $13.19\pm0.08$  & $20000\pm1000$ \\ 
          &   59760.08 & F2   & 1700-3200 & 2.00--2.32 & 1.03&1.03&   $''$     &  &                    \\ 
July 5    &   59765.10 & F2   & 200-1200  & 0.9--2.5 &1.19&1.19 &  28.38 & $13.86\pm0.07$  & ---  \\
          &   59765.10 & F2   & 1700-3200 & 2.00--2.32 & 1.27&1.32   &  $''$     &         &         \\
July 16   &   59776.27 & GNIRS& 1200  & 0.9--2.5  &1.29& 1.26 & 39.55 & $15.62\pm0.14$  & ---   \\
July 22   &   59782.32 & GNIRS& 1200  & 0.9--2.5 &1.34&1.32  & 45.60 & $15.48\pm0.07$  & ---         \\\hline
\multicolumn{10}{l}{$^*$See http://www.gemini.edu/instrumentation/flamingos-2/components\#Grisms
for how $R$ for F2 varies across the spectral interval.}\\
\multicolumn{10}{l}{$^\dag~t=0$ was at MJD 59736.72.}
\end{tabular}
\end{table*}

\us\ was reported to be in outburst on 2022 June 06.72 \citep[MJD 59736.72;][]{moriyama22}, 
which we take as the time origin. The visual light curve from 
the AAVSO archive\footnote{https://www.aavso.org/} is shown in Fig~\ref{LC}. 

The 2022 eruption has been extensively observed at radio \citep{sokolovsky22},
optical \citep{siviero22,woodward22} and X-ray \citep{orio22,page22a,page22b} 
wavelengths. 
{Preliminary descriptions of NIR observations were given
by \cite{banerjee22} and \cite{rudy23}.}
We present here the complete series of our NIR spectroscopic observations.

\section{Observations}
\label{obs}

\begin{figure*}
 \includegraphics[width=15cm,keepaspectratio]{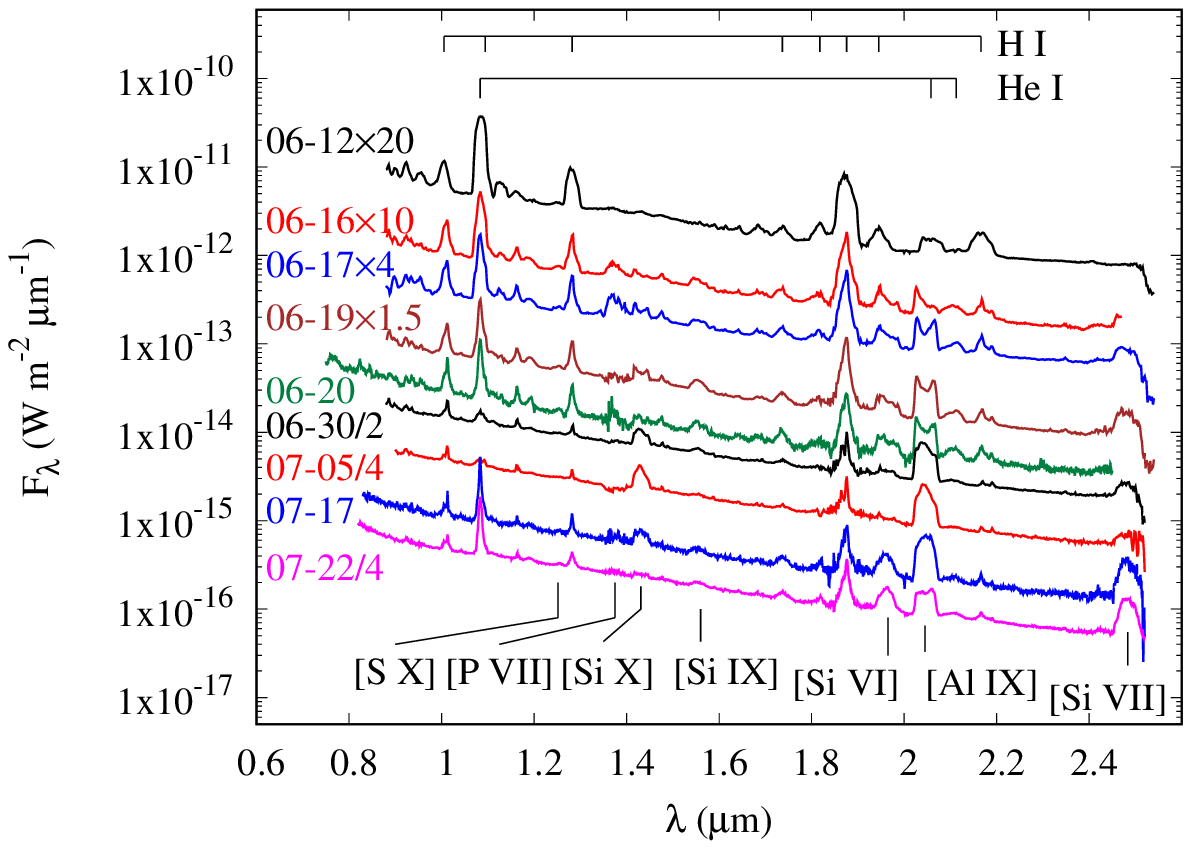}
 \includegraphics[width=12cm,keepaspectratio]{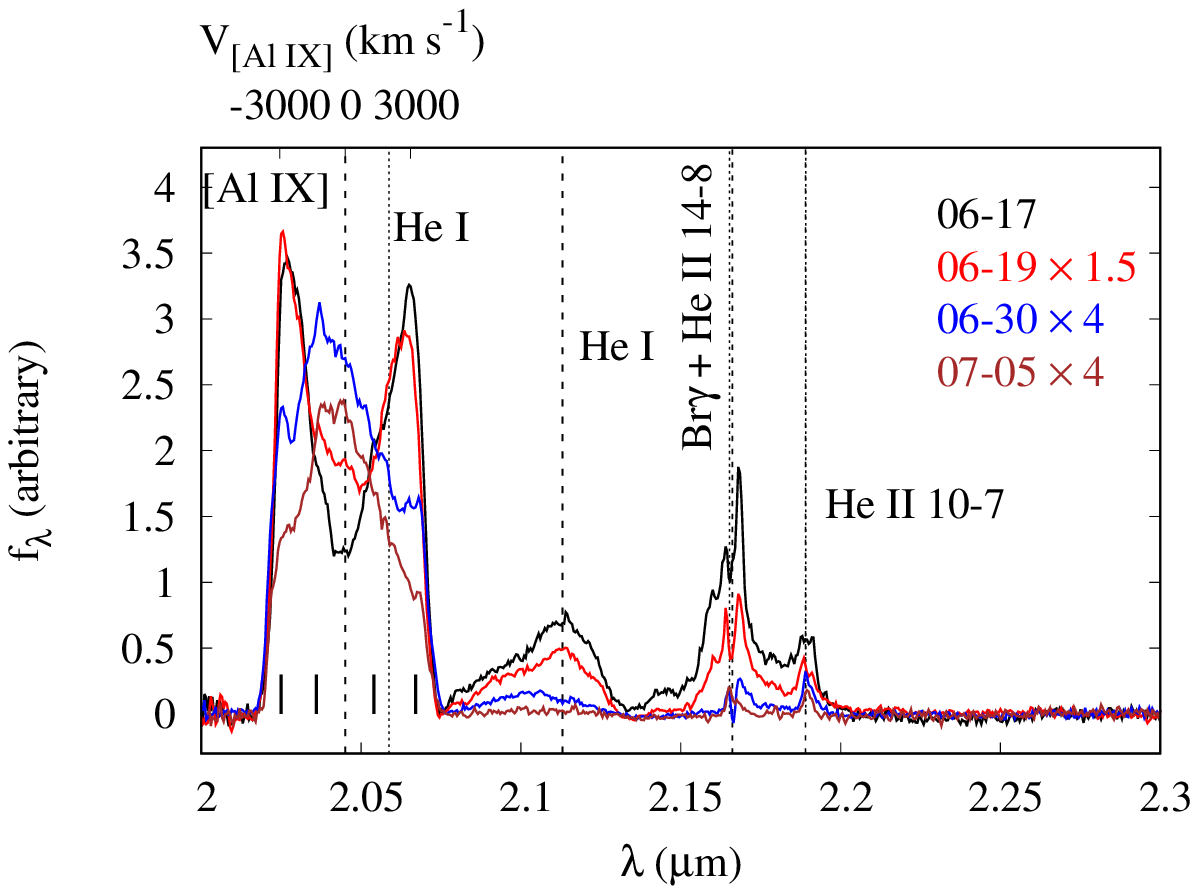}
 \caption{Top: sequence of lower resolution spectra. The 2022
 spectra are labelled by date
 in the form MM-DD, and have been multiplied/divided by the 
 factors indicated. Prominent \pion{H}{i}, \pion{He}{i} and coronal lines
 are identified. 
 Bottom: R3k spectra, labelled by date in the form MM-DD, 
 have been multiplied/divided by the  factors indicated.
 Top radial velocity scale (in \vunit) is for \fion{Al}{ix} only.
 {Black lines at bottom of panel indicate 
 approximate wavelengths of the 
  ``core'' and ``horn'' components.}
 See text for details.\label{data}}
\end{figure*}

\subsection{Gemini Observatory}

Infrared spectra covering all, or portions, of the 0.9--2.5\mic\ wavelength
interval were obtained at the Gemini South and North telescopes from 
2022 June 12 to 2022 July 22 using the facility instruments FLAMINGOS-2
\citep[F2;][]{eikenberry04} and the Gemini 
Near-InfraRed Spectrometer \citep[GNIRS;][]{elias06}.
Observing details are provided in Table~\ref{log}.  
Most of the spectra were obtained with the $J\!H$ and $H\!K$ grisms in F2, each
of which provides resolving powers, $R$, ranging from 200 to 1400 
(corresponding to 200--1500\vunit). Several higher resolution 
spectra of a large portion of the $K$ band were obtained with F2's 
R3k grism.  GNIRS in its cross-dispersed mode and its 32$\,\ell\,${mm}$^{-1}$ 
grating was employed for the final two epochs. Observing times 
varied from one minute during the early epochs when \us\ was bright,
to 32 minutes for the final epoch when it had faded considerably.  

A0 dwarfs served as telluric standards. They were observed at air masses 
closely matching  those of \us, either immediately before, or immediately 
after, \us. Data reduction utilizing both IRAF \citep{tody86,tody93}
and FIGARO \citep{currie14} employed the standard procedures of 
spectrum extraction, spike removal, wavelength calibration (using spectra 
of argon arc lamps), removal of \pion{H}{i} lines from the spectra of 
the standard star, cross-correlating the spectra of \us\ and the 
telluric standard star, shifting the spectrum of the former to 
align it with the spectra of the standard star, and ratioing 
the former by the latter. 

\subsection{IRTF}

NASA Infrared Telescope Facility (IRTF) observations were obtained with 
SpeX \citep{rayner03} in short-crossed dispersed (SXD) mode with
a $0\farcs5 \times 15''$ slit (spectral resolving power 1200,
corresponding to $\simeq 250$\vunit), in sub-arcsec seeing 
(FWHM [$K$] $\approx 0\farcs65$) under photometric conditions. The source 
was nodded between two positions along  the  slit.
These data (comprising 478~s of total on-source integration time) were 
reduced and corrected for telluric absorption(s), flux calibrated, and order 
merged with the SpexTool pipeline \citep{cushing04}. The A0V standard star 
(for telluric correction) HD 148968 was observed at comparable air mass.
The accuracy of flux calibration of the resultant spectra is $\simeq10$\%.

Our time coverage is {one of} the longest over 
which an eruption of \us\ has been
followed in the NIR.
A difficulty with following this nova for longer periods is
its very rapid fading, the time $t_2$ ($t_3$) to decline by 2 (3) 
magnitudes from maximum, being 1.2 (2.6) days \citep{schaefer10}.

Where possible, aperture photometry was carried out on acquisition frames 
obtained in the $H$ band to facilitate positioning of the object in the
spectrograph slit. Calibration was achieved using multiple field stars with
known $H$ magnitudes from the 2MASS \citep{2mass} 
and UKIDSS \citep{dye06} surveys. These $H$ 
magnitudes are listed in Table~\ref{log},
and included in Fig.~\ref{LC}. This figure also includes the $H$-band light 
curve from the 2010 eruption \citep[][see that paper for details]{pagnota15}, 
on which our $H$ band 
magnitudes are superposed. The 2022 $H$ photometry agrees well with 
the 2010 $H$ light curve, and there is excellent agreement between the 
shapes of the 2010 $H$ and 2022 $V$ light curves.

\section{Overview of the spectra}

The lower resolution ($\leq1200$) spectra are shown in Fig.~\ref{data} 
(top panel). \pion{H}{i} recombination lines are prominent, but they are not
useful for analysis, because they coincide with
\pion{He}{ii} recombination lines. For example, six \pion{He}{ii}
lines are close in wavelength to Br\,$\gamma$ at 2.1661\mic.
The strongest of these is the 10--7 transition 
at 2.1891\mic, while \pion{He}{ii} 14--8 (2.1652\mic) almost exactly 
coincides with Br\,$\gamma$ (see Fig.~\ref{data}, bottom panel).

Low excitation lines are often seen in the IR spectra of novae
in eruption \citep[see compilation in][]{banerjee12a}. 
However, other than the \pion{H}{i} recombination lines,
there were no low excitation lines in the IR spectra of \us.
\pion{O}{i} 1.1287\mic\ was present up to day~10.42, but the upper
level of this transition is pumped by Ly$\,\beta$ fluorescence, 
thus pointing to a high excitation environment.

\subsection{Coronal lines\label{coronals}}

\begin{figure*}
   \includegraphics[width=8cm,keepaspectratio]{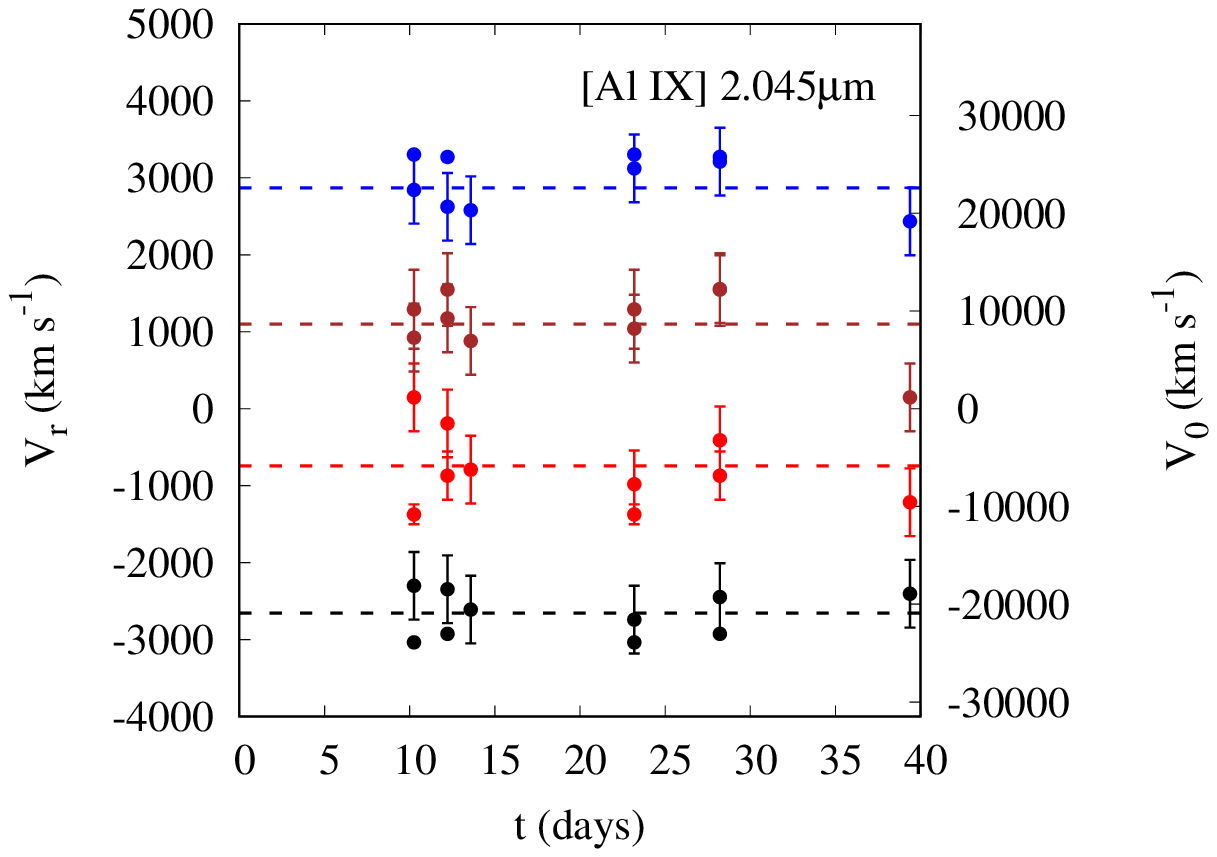}
  \includegraphics[width=8cm,keepaspectratio]{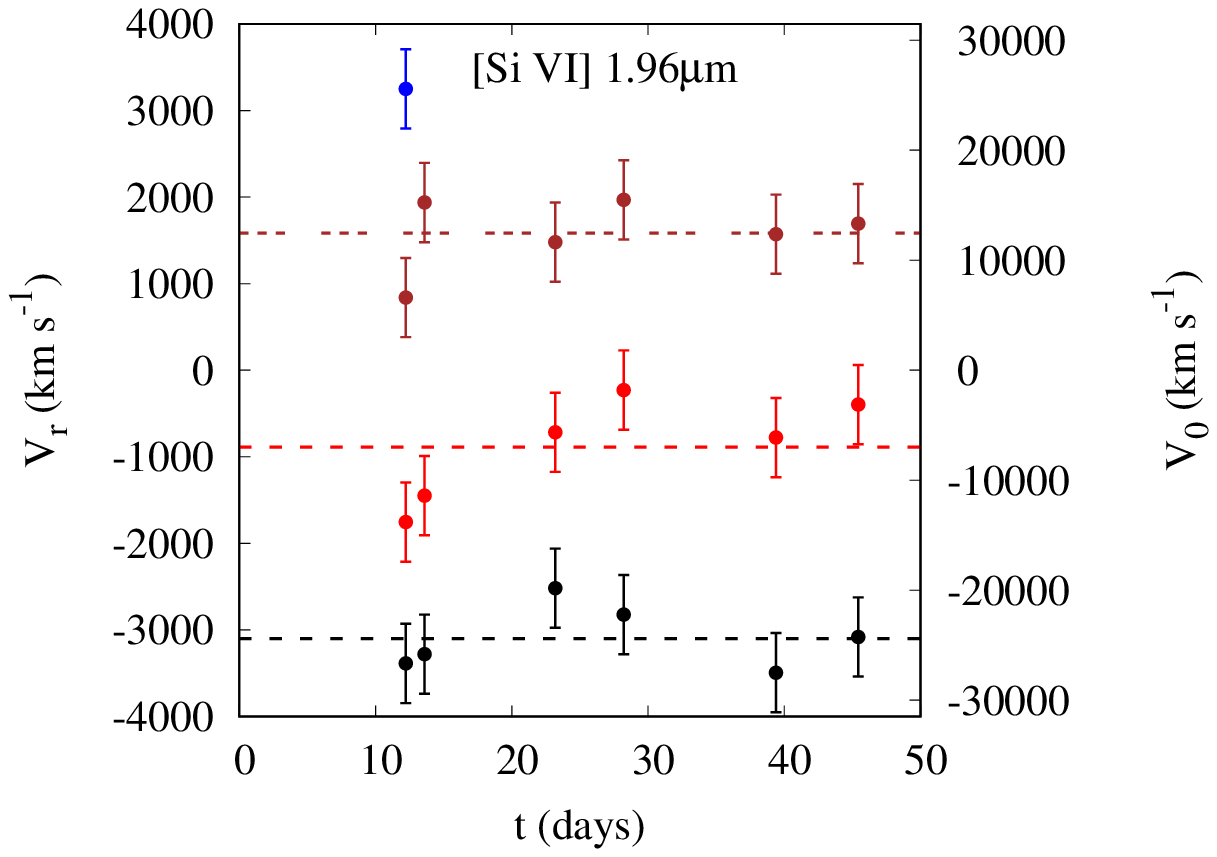}
  \caption{Left: velocities as determined from the \fion{Al}{ix} line;
  left axis, radial velocity $V_{\rm r}$, right axis, velocity $V_0$
after allowing for  binary inclination. 
{Red points and lines, core components; blue/black lines
and points, horn components.}
  Right: velocities as determined from the \fion{Si}{vi} line;
  left axis, radial velocity, right axis, velocity after allowing for 
  binary inclination. The blue point is the fourth component, 
  seen at day 9.41 only. 
  {Red points and lines, core components; black line
and points, horn components.}
  In both panels dashed lines are mean radial velocities.
  \label{coronal2}}
\end{figure*}

A preliminary account of our data was given by \cite{banerjee22},
who reported the first identification of coronal lines in an outburst
of \us. Seven coronal lines: \fion{S}{ix} 1.252\mic, 
\fion{P}{vii} 1.375\mic, \fion{Si}{x} 1.431\mic, 1.560\mic, 
\fion{Si}{vi} 1.965\mic, \fion{Al}{ix} 2.045\mic\ and 
\fion{Si}{vii} 2.483\mic\ were clearly present on 
day~9.41 (see Fig.~\ref{data}). The last of these is at the extreme 
red end of the spectrum  
{(where telluric correction is challenging)}, 
which limits its usefulness as a diagnostic. This is 
{one of the earliest detection of coronal lines in any nova, 
classical or recurrent}. 
\cite{rudy23} did not detect any of these coronal
lines in a spectrum obtained only 20~hours earlier.
We did not detect the \fion{Ca}{viii} line at 2.321\mic, which
is often seen in the coronal phase of novae.

\begin{figure*}
  \includegraphics[width=8cm,keepaspectratio]{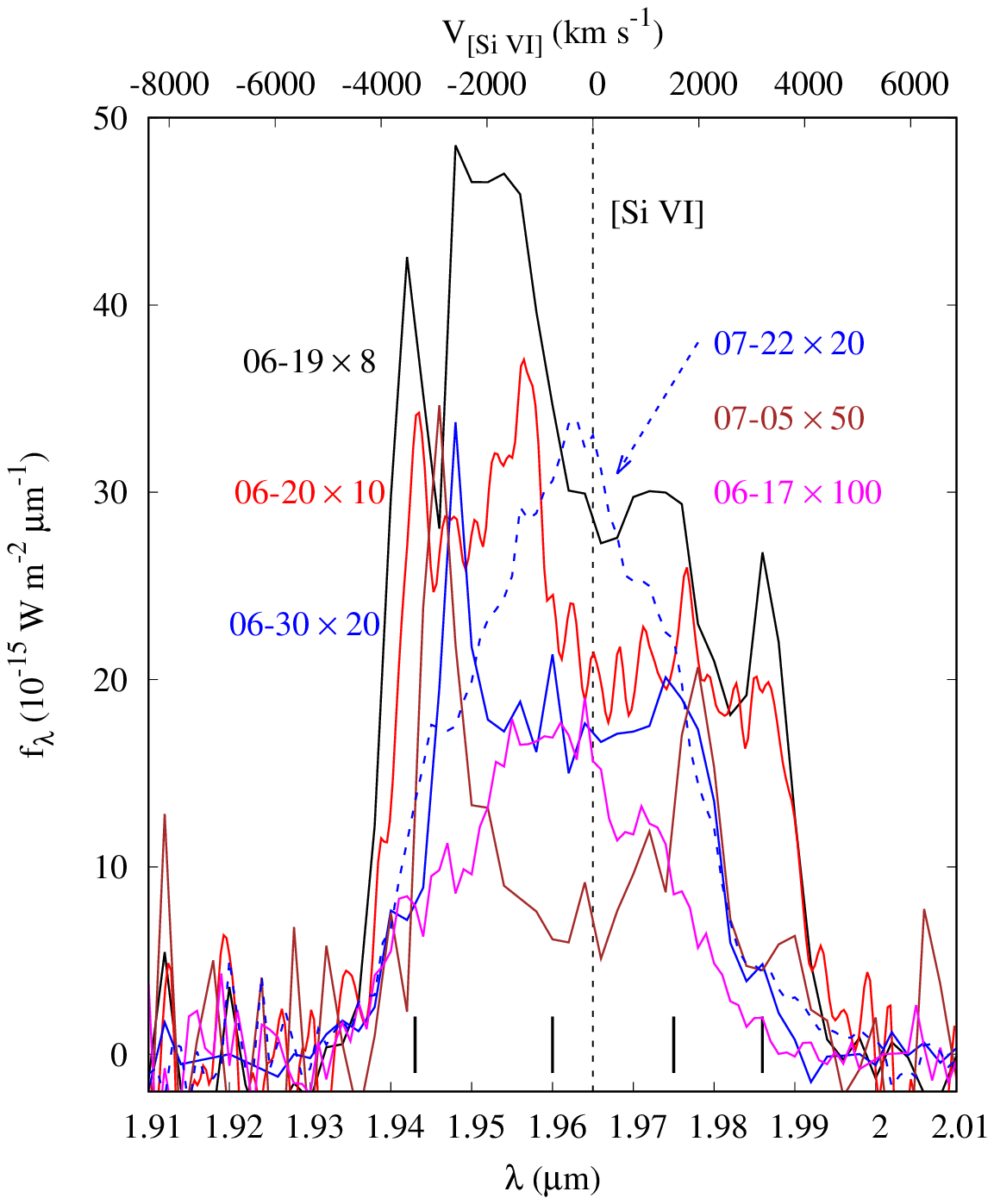}
  \includegraphics[width=8cm,keepaspectratio]{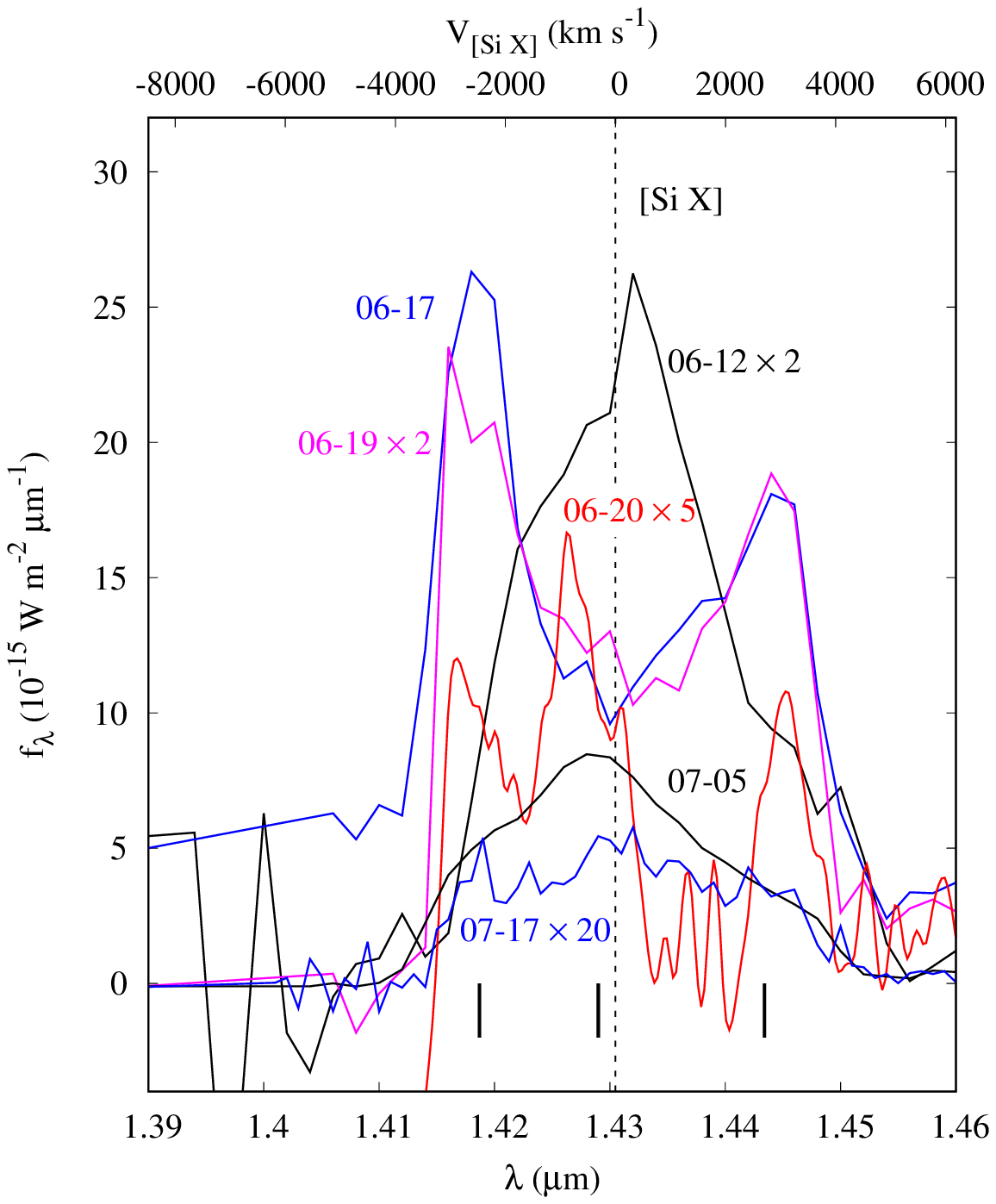}
  \caption{{Left: Evolution of the \fion{Si}{vi} 1.965\mic\ profile
  as revealed by the low resolution spectroscopy; 
  dates are given in the form MM-DD, and spectra 
  have been multiplied by the values indicated.
  Right, as left, for selection of \fion{Si}{x} 1.431\mic\ profiles.
  In each case the top scale gives the radial velocity, and
  the black lines at bottom of panels indicate approximate wavelengths of the 
  ``core'' and ``horn'' components.}
   \label{coronal1}}
\end{figure*}

In the spectra of 16 June and later, we consider that the feature
at $\sim2.05$\mic\ is mainly due to \fion{Al}{ix} $\lambda=2.045$\mic.
\pion{He}{i} $\lambda=2.059$\mic\ contributed partially 
on that date, but faded in subsequent spectra.
This feature was present in both the low and high resolution data. 
In all cases the line profile is complex, 
and can (in common with several other lines) be divided into four
components. The evolution of the \fion{Al}{ix} line, as revealed by the 
high-resolution data, is shown in Fig~\ref{data}. Its profile on June 
17 and 19 has a characteristic ``horned'' structure, typical of 
jets\footnote{By ``jets'' here we mean highly collimated outflows.} 
emitted orthogonal to the binary plane \citep[][]{menzel33,kato03},
together with two lower velocity components (the ``core'') 
that likely arise in the 
binary plane. In time the horns decline in flux, while the core 
increases relative to the horns. 
There is no evidence that the velocities, as determined from the peak
wavelengths of the lines, changed with time.
The mean radial velocities, $V_{\rm r}$, for the four components are
$+2870\pm60$\vunit, $+1000\pm130$\vunit, $-770\pm130$\vunit, and 
$-2660\pm80$\vunit\ (see Fig~\ref{coronal2}, left).
Assuming that the horns do indeed arise in jets ejected perpendicular to the 
binary plane, the velocities, $V_0=V_{\rm r}/\cos{i}$,
implied by these components, allowing for the binary inclination of 
$82\fdg7$, are $+22600\pm1000$\vunit, and $-20900\pm600$\vunit\
(see Fig.~\ref{coronal2}, left).

The above uncertainties are the standard errors in the mean 
radial velocities; including the uncertainty in
the inclination ($\pm2\fdg9$) gives $+22600^{+14800}_{-6400}$\vunit, 
and $-20900^{+13700}_{-5900}$\vunit\ respectively. However
as \us\ is an eclipsing system, the constraint that
the binary inclination $i>\cos^{-1}(R_*/a)$, where $a$ is the binary
separation \citep[6.5\Rsun;][]{thoroughgood01} must be included, 
so that $i>80^\circ$ for a K2IV secondary. This means that the 
lowest value allowed by these uncertainties is a physical lower limit. 
Remarkably, such high velocities are comparable to those measured 
in supernovae in the NIR \citep[e.g.,][]{hsiao19}.

A further, and compelling reason, to associate the horns with a fast 
polar flow, orthogonal to the binary orbital plane that contains a 
material overdensity \citep{drake10}, is that they are seen only 
in the coronal lines, not in the \pion{H}{i} lines (see Fig.~\ref{data}). 
Assuming geometric dilution of the electron density $n_e\propto{r}^{-2}$,
the coronal lines are expected to be first seen from regions that have 
dispersed the most rapidly, allowing the electron density to fall 
below the critical density for each transition. The kinematics of a 
bipolar flow, which is collimated by an equatorial overdensity, 
are such that matter along the polar direction always has the highest 
velocity, and thus declines most rapidly in density, allowing the 
emergence of coronal lines. At later times, emission in the coronal 
lines may be expected to be 
seen from other regions of the ejecta (e.g., matter in the equatorial region),
as it is expanding at lower velocities and will thus take more time to 
drop below the critical density. The \pion{H}{i} lines are likley to have
been emitted from all regions of the ejecta. 
Horned profiles were also seen in the coronal lines, 
but not in the H and He lines, in the classical nova V5668~Sgr 
\citep[see Figure 16 of][]{gehrz18}, who also argued that they arise in
fast flowing polar material.

In our early spectra of the 2022 eruption, 
the \fion{Si}{vi} 1.965\mic\ line (which 
was observed at low resolution only) is overwhelmed by Br$\,\delta$ at
1.9451\mic, but as the latter fades the \fion{Si}{vi} line predominates. 
Its profile over the period 
{June~17} to July~22 is shown in
Fig.~\ref{coronal1}. 
The mean implied velocities of the horned components, allowing for the 
binary inclination, are $-24400\pm1200$\vunit\ and $12500\pm1300$\vunit,
where the uncertainties here are the standard errors of the means.
Including the uncertainty in the binary inclination gives
$-24400^{+16000}_{-6900}$\vunit\ and $12500^{+8200}_{-3500}$\vunit\
respectively (see Fig.~\ref{coronal2}, right).

The temperature of the coronal gas can be determined
using the iterative method outlined in \cite{woodward21} and 
\cite{evans22}. We assume $E(B-V)=0.2\pm0.1$
\citep[cf.][]{schaefer10}, and use effective collision strengths 
$\Upsilon$ (i.e., the collision strength $\Omega$ averaged over 
a thermal electron energy distribution) from the Iron Project
\citep{hummer93,lennon94,galavis97,badnell06} online 
database\footnote{http://cdsweb.u-strasbg.fr/tipbase/home.html},
as well from the literature \citep{brage00}.
The $\Upsilon$ values used are listed in Table~\ref{crit}.

In view of the complexities of the line profiles, 
one can not simply take ratios of total line fluxes, but must identify 
individual velocity components. The components most easily isolated
are those that have the highest radial velocities,
namely the ``+2870\vunit'' and ``-2660\vunit'' components (i.e., the
horns/jets); {we refer to these as the ``red~jet'' 
and ``blue~jet'' components respectively.}
We use the \fion{Si}{vi} $\lambda=1.965$\mic\ and
\fion{Si}{x} 1.403\mic\ lines. The Si coronal line (\fion{Si}{ix}
$\lambda=1.560$\mic) is problematic as it is superimposed on higher order
members of the hydrogen Brackett series. 

The coronal temperatures $T_{\rm cor}$ for these components are 
listed in Table~\ref{cor-temp}. There is no evidence for any 
temporal change in $T_{\rm cor}$ for either component, the mean values being
($T_{\rm cor}$ in K)
$\log{T_{\rm cor}}=5.88\pm0.04$ for the 
{blue jet} component, and
$\log{T_{\rm cor}}=5.82\pm0.04$ for the 
{red jet} component;
the overall mean is $\log{T_{\rm cor}}=5.85\pm0.03$ 
($T_{\rm cor}=7.1[\pm0.5]\times10^5$~K, $kT=61.3$~eV), 
which we use henceforth.

The critical electron densities (above which the upper levels are
mainly collisionally rather than radiatively de-excited), are given 
in Table~\ref{crit}. They have been
calculated at $\log{T_{\rm cor}}=5.85$, except for \fion{S}{ix}
and \fion{Si}{vii}, for which $10^5$~K is the highest value available 
in the Iron Project database.

\begin{table}
\caption{Coronal gas temperatures for the ``jets'',
determined using Si lines as described in text.\label{cor-temp}}
 \begin{tabular}{cccc}\hline
  Date & Day & \multicolumn{2}{c}{$\log{T}$ (K)} \\ \cline{3-4}\noalign{\smallskip}
       &     & Blue jet  & Red jet \\ \hline
       June 16 & 9.41 & 6.03 & 6.01 \\
       June 17 & 10.42 & 5.90 & 5.80     \\
       June 19 & 12.39 & 5.82 & 5.76 \\
  June 30 & 23.36 & 6.02 & 5.90   \\
  July 5 & 28.38 & 5.77 & 5.70 \\
  July 16 & 39.55& 5.75 & 5.76 \\
  July 22 & 45.60 & 5.90 & 5.83\\
  \hline
 \end{tabular}
\end{table}

\begin{table}
\caption{Critical densities at $7.1\times10^5$~K for coronal lines. 
IP is the ionisation potential of the lower ionsation state.
$\Upsilon$ is the effective collision strength at $7.1\times10^5$~K
unless otherwise noted.
See text for details, and for definition of $T_*$. \label{crit}}
 \begin{tabular}{cccccc}\hline
 Line & $\lambda$ (\mic) & IP (eV) & $kT_*$ (eV) & $\Upsilon$ & \ncrit (cm$^{-3}$) \\\hline\noalign{\smallskip}
\fion{S}{ix} & 1.252  & 329 & 137 & 0.907$^*$ & $6.30\times10^9$ \\
\fion{P}{vii} & 1.375 & 220 &  93 & 0.315 & $4.30\times10^9$ \\ 
\fion{Si}{x} & 1.431  & 351 & 146 & 0.415 & $1.01\times10^9$ \\
\fion{Si}{ix} & 1.560  & 304 & 127 & 0.275$^*$ & $1.02\times10^9$\\
\fion{Si}{vi} & 1.965 & 167 &  70 & 0.404 & $2.32\times10^9$ \\
\fion{Al}{ix} & 2.045 & 285 & 119 & 0.720 & $2.94\times10^8$ \\
\fion{Si}{vii} & 2.483 &205 &   86 & 0.695$^*$ & $1.04\times10^9$ \\   \hline
\multicolumn{6}{l}{$*$At $10^5$~K.} \\
 \end{tabular}
\end{table}

\subsection{The \pion{He}{i} lines}
\label{hei}
\begin{figure}
 \includegraphics[width=8cm,keepaspectratio]{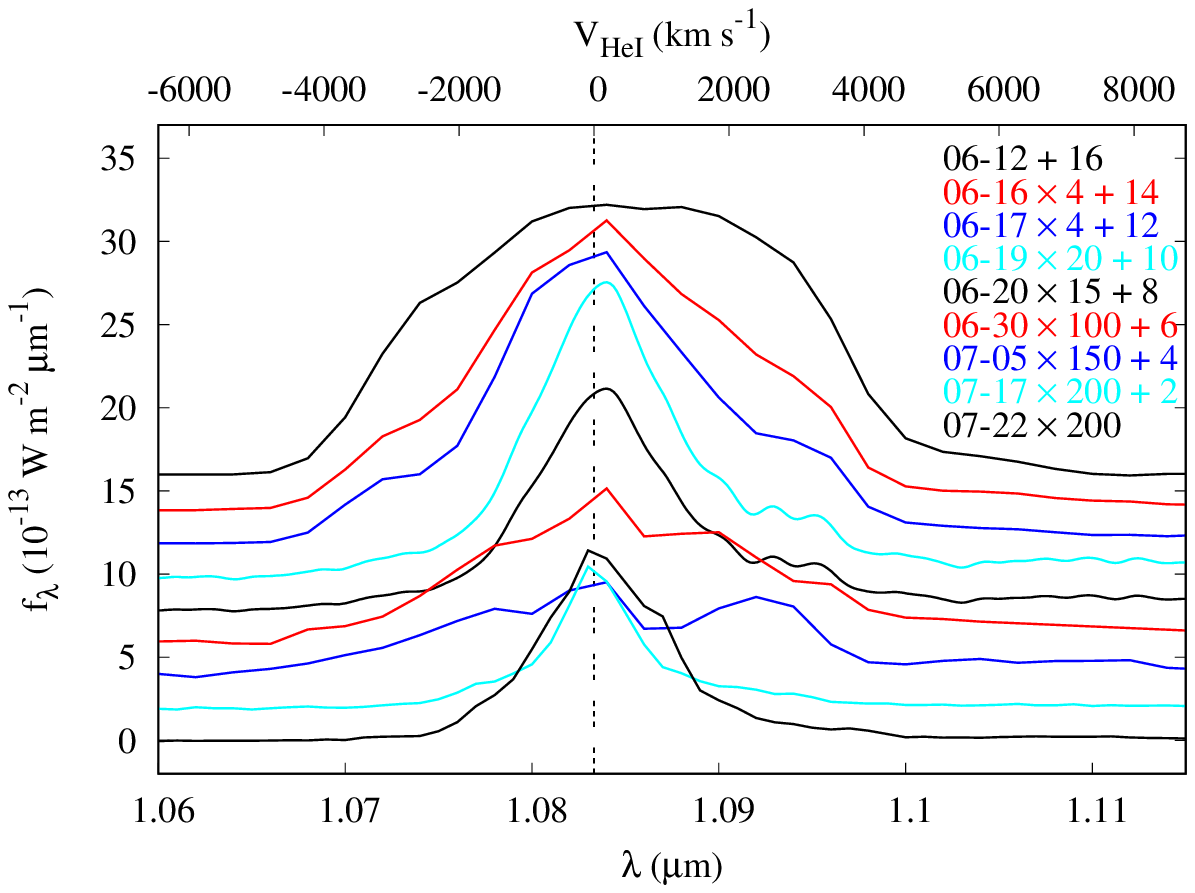}
  \includegraphics[width=9cm,keepaspectratio]{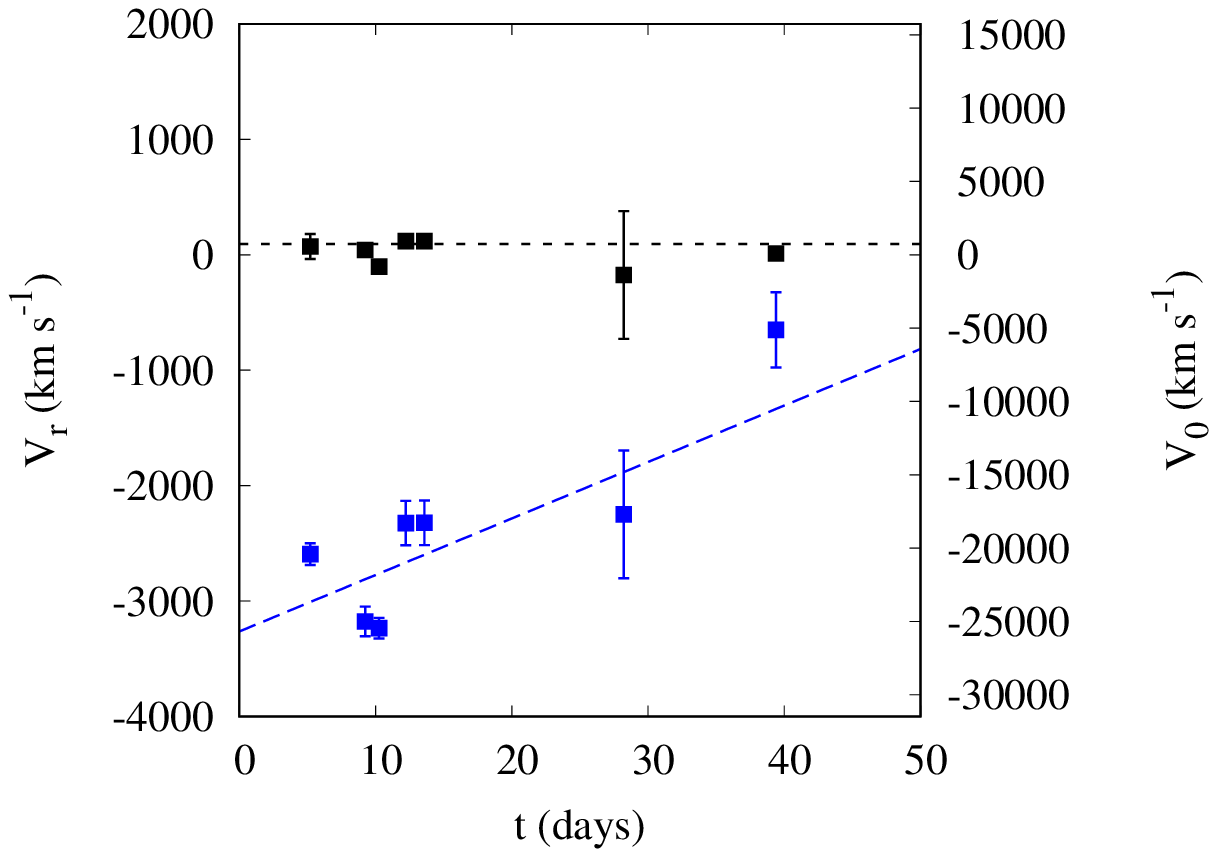}
  \caption{Top: evolution of the \pion{He}{i} 1.083\mic\ profile
  as revealed by low resolution spectroscopy. Spectra have been multiplied
  and displaced by the factors indicated. Dates (2022) are given 
  in the form MM-DD.
   Bottom: velocities of the core (filled black squares) and blue 
   (filled blue squares) wing components of \pion{He}{i} 1.083\mic. 
The left-hand axis gives the radial velocity $V_{\rm r}$; the right-hand axis 
gives the space velocity $V_0$ (after allowing for binary inclination)
giving rise to the blue wing. 
Black dotted line is the mean velocity as determined from the 
peak of the \pion{He}{i} 1.0833\mic\ core; blue dotted line is a
linear fit to the \pion{He}{i} 1.0833\mic\ wing. 
\label{evol}}
\end{figure}

\begin{figure}
 \includegraphics[width=8cm,keepaspectratio]{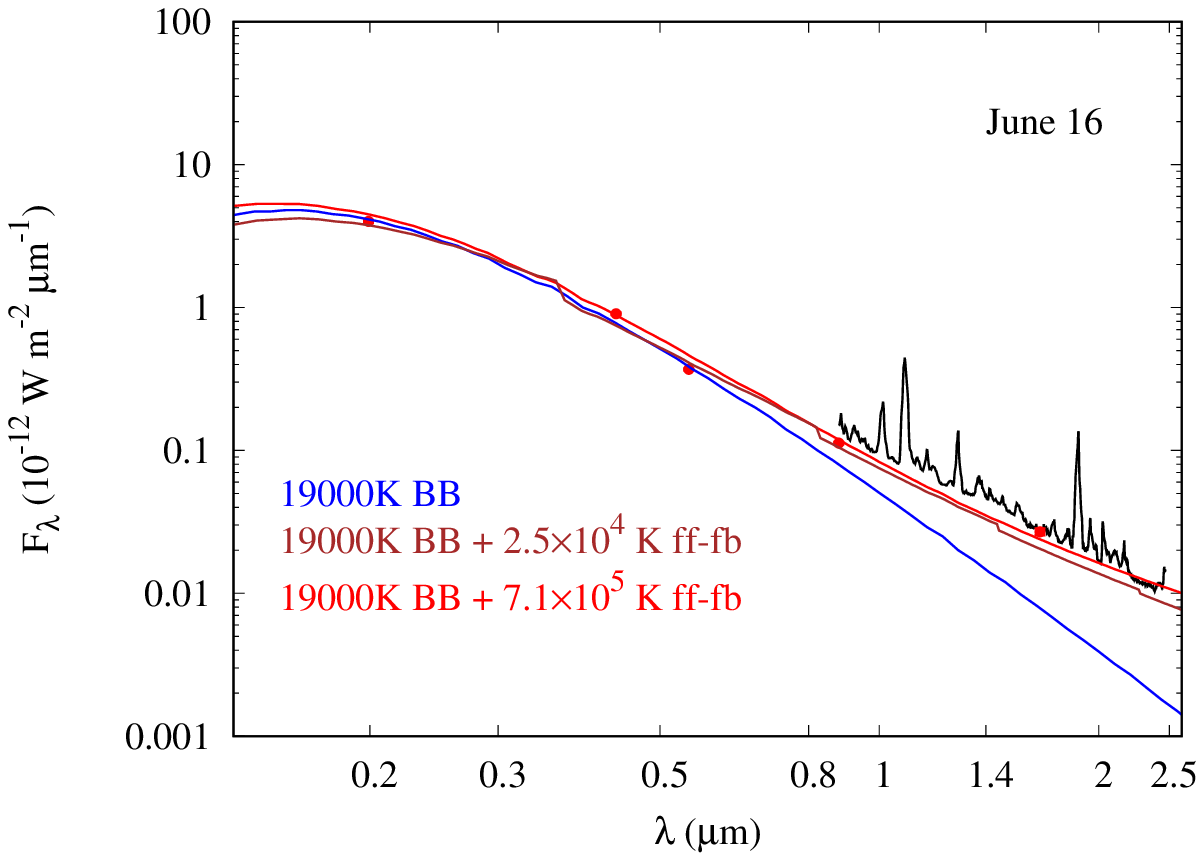}
 \includegraphics[width=8cm,keepaspectratio]{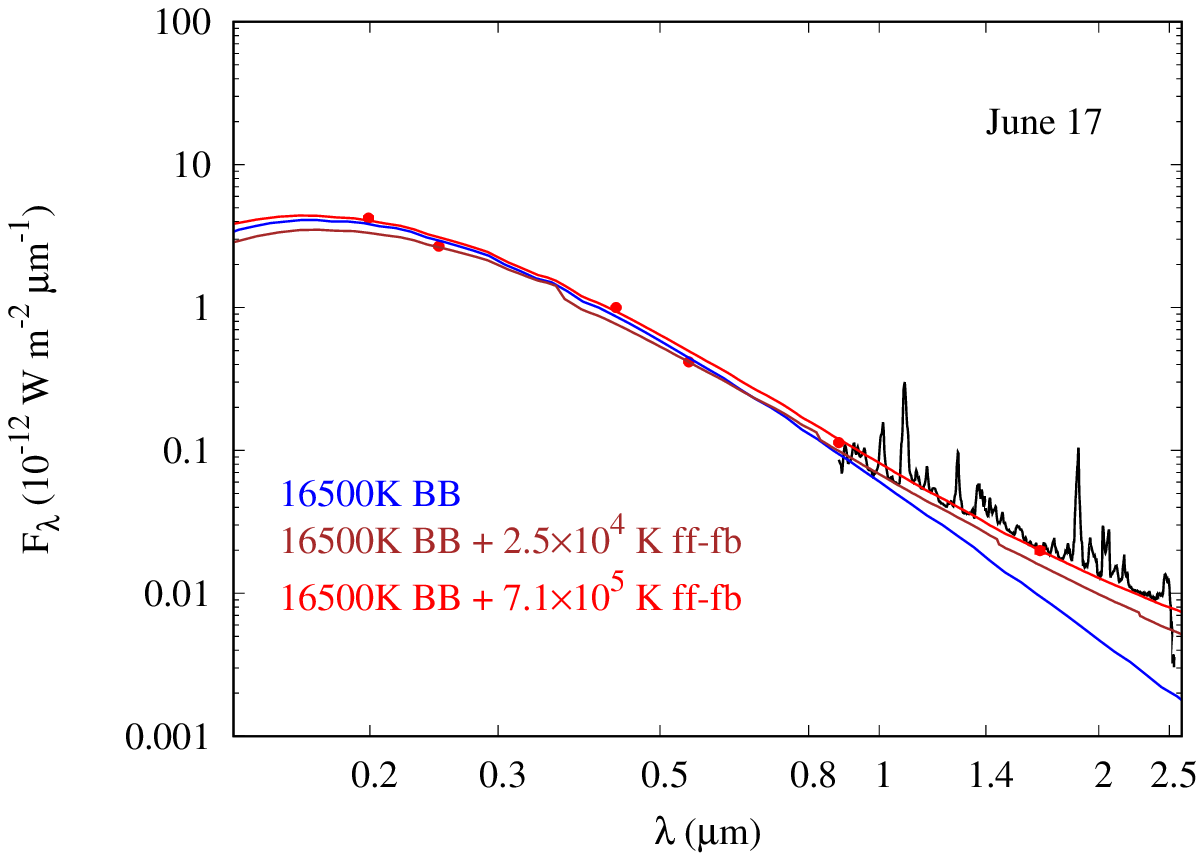}
 \includegraphics[width=8cm,keepaspectratio]{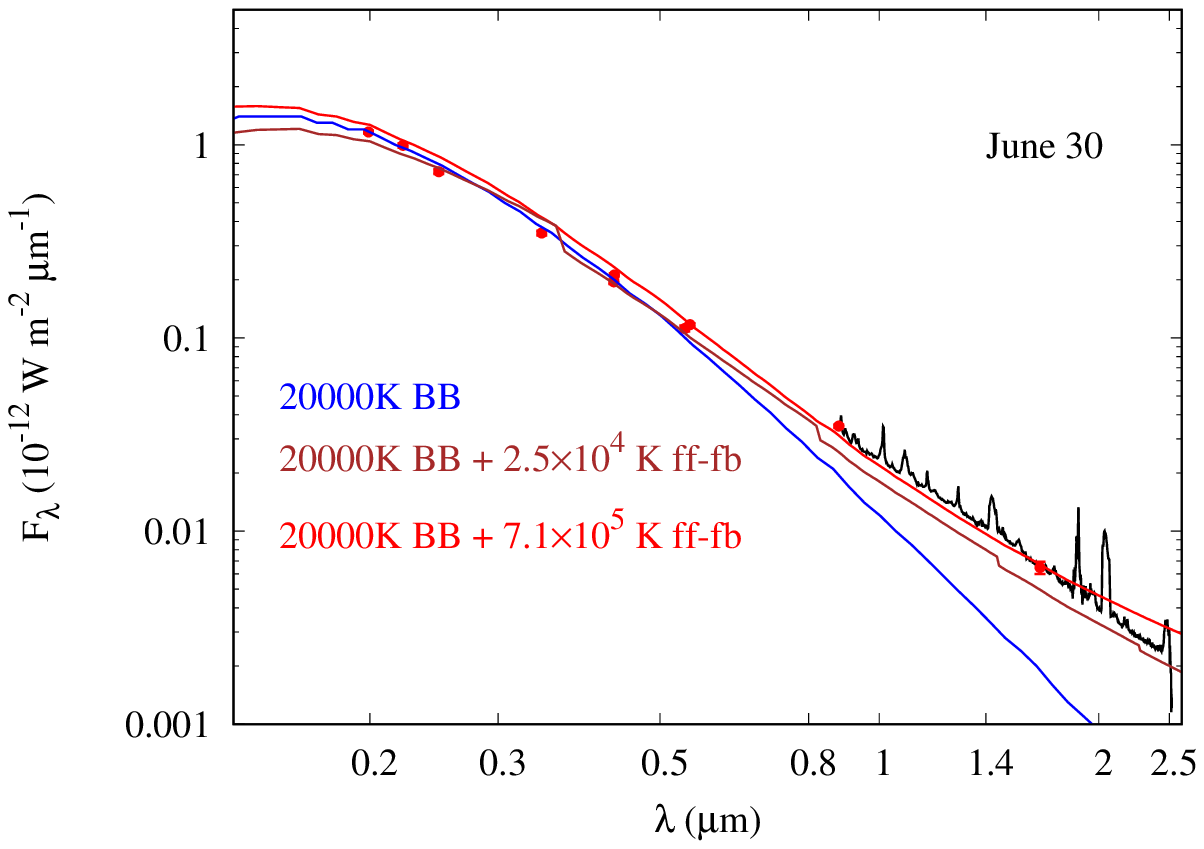}
 \caption{Top: UV-optical-NIR SED of \us\ on June 16. 
 Data dereddened by $E(B-V)=0.2$.
 Red points: Swift UVOT photometry and $H$ band photometry obtained
 as described in Section~\ref{obs}. 
 Blue curve is black body fit to UVOT data. Red curve is black body +
 free-free/free-bound emission at $7.1\times10^5$~K; brown curve is black body +
 free-free/free-bound emission at $2.5\times10^5$~K. 
 Middle: as top, but for June 17.
 Bottom: as top, but for June 30.
 In all cases, helium is entirely neutral for the $2.5\times10^4$~K
 case, and completely ionised for the $7.1\times10^5$~K
 See text for details. \label{SED}}
\end{figure}

\begin{figure}
 \includegraphics[width=8cm,keepaspectratio]{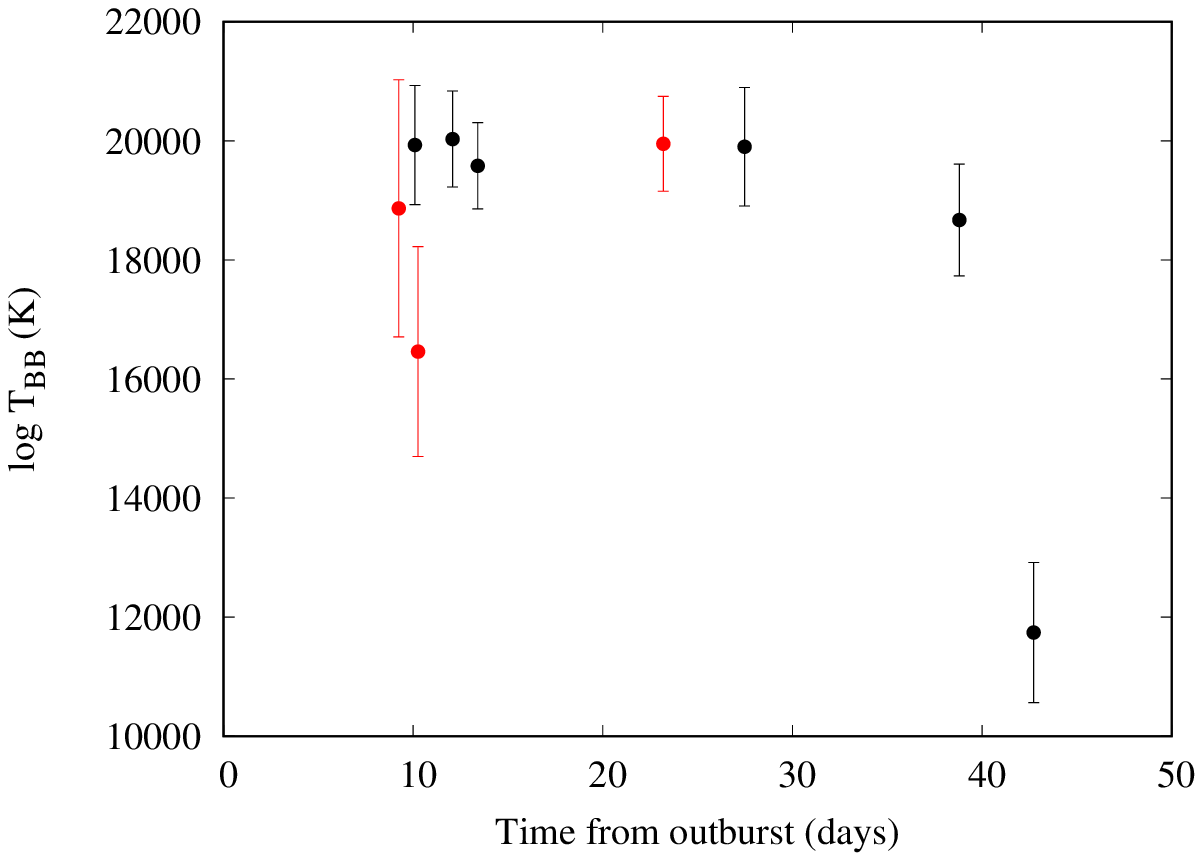}
 \includegraphics[width=9cm,keepaspectratio]{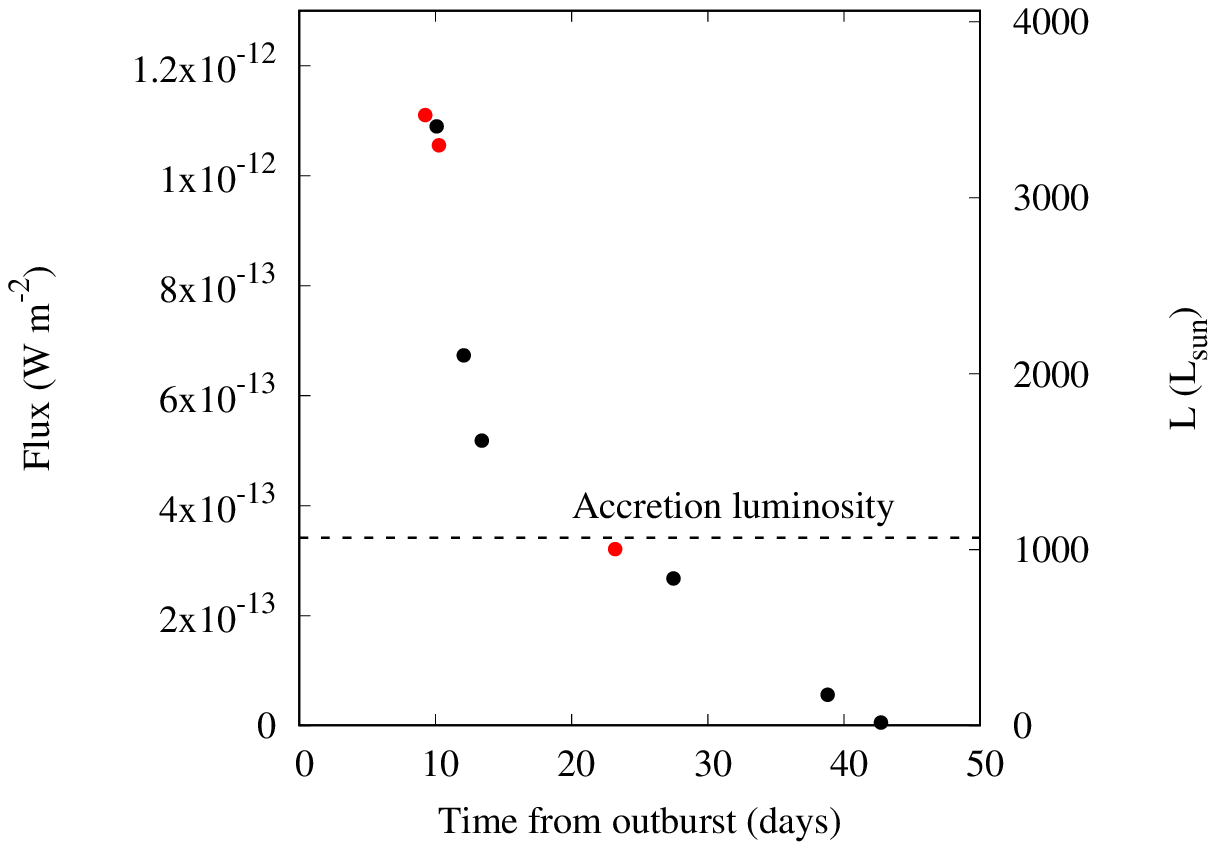}
 \includegraphics[width=9cm,keepaspectratio]{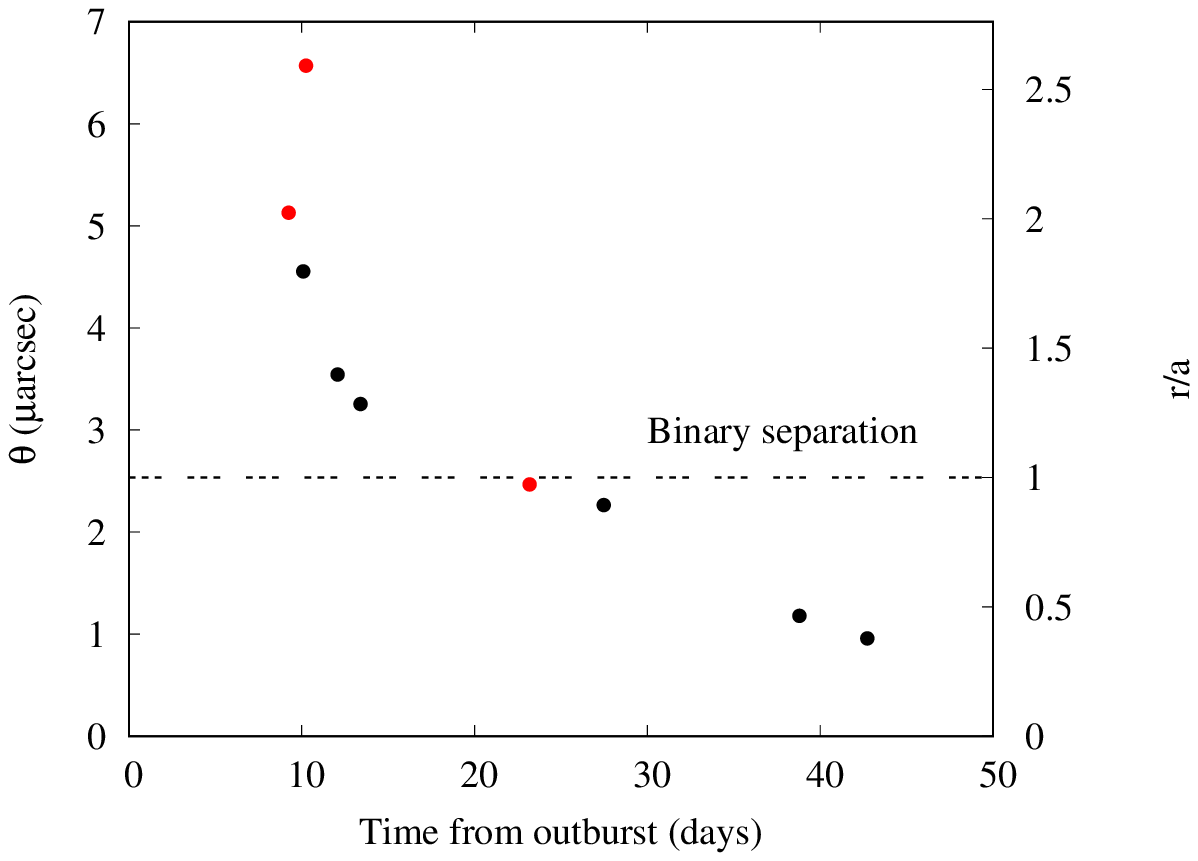}
 \caption{Top: time-dependence of $T_{\rm BB}$.
 Middle: time-dependence of black body flux (left axis),
 and luminosity, assuming 10~kpc distance (right axis).
 Bottom: time-dependence of black body angular diameter (in $\mu$\,arcsec,
 left axis) and diameter (in units of binary separation, right axis) of black body.
 Distance assumed is 10~kpc; binary separation assumed is 6.5\Rsun\
 \citep{thoroughgood01}.
 Black points, black body fits to Swift data; red points, 
 black body fits to Swift data within 88~min of IR observations.
 \label{BB}}
\end{figure} 

On June 12 (day~5.38), the \pion{He}{i} 1.083\mic\ line
was broad, with at least four components.
Thereafter, it consisted of a strong core, with wings to the blue and red
(see Fig.~\ref{evol}, top panel). In each spectrum there is almost certainly a 
contribution from Pa$\,\gamma$ 1.094\mic, which 
is superimposed on the contribution 
from any red wing the \pion{He}{i} line may have. 
For plasma temperatures up to $3\times10^4$~K, the Pa$\,\gamma$ 
flux is expected to be $\sim0.6$ that of Pa$\,\beta$ for a wide 
range of electron densities if ``Case~B'' applies. Thus, for 
example, on June 16 Pa$\,\gamma$ would have had a peak flux 
above the adjacent continuum of 
$\sim9\times10^{-14}$~W~m$^{-2}\mic^{-1}$,
comparable with $\sim7\times10^{-14}$~W~m$^{-2}\mic^{-1}$ for the 
peak of the red wing of the observed feature, and much less than the
$4.1\times10^{-13}$~W~m$^{-2}\mic^{-1}$ for the peak of
\pion{He}{i} itself. Therefore, in our analysis below,
we consider the blue wing only.

The central wavelengths of the core and blue wings have 
been converted to radial velocities, as shown in Fig.~\ref{evol}. 
There is no evidence that the radial velocity of the material 
responsible for the core component changes with time.
The implied radial velocity is essentially
zero (mean value $=95\pm120$\vunit). The mean FWHM 
of the core emission is $750\pm100$\vunit.
The wavelength of peak flux of
the core is close to zero; thus
we conclude that this component arises in material (e.g., a torus)
ejected in the orbital plane, almost perpendicular 
to the plane of the sky. 

On the other hand there is some evidence that the material emitting 
the blue wing decelerated as
\[ V_{\rm r} (\vunit)= -3260 [\pm280] + 49 [\pm 11] \:\: t \]
where $t$ is in days. The implied deceleration is 
considerably less than that implied by the narrowing of
H$\,\alpha$ during the 2010 eruption \citep{munari99}, but is
of the same order as the deceleration implied by the narrowing of 
Pa$\,\beta$ in the 2019 eruption of V3890~Sgr \citep{evans22}. 
The initial radial velocity of the blue wing ($\simeq-3000$\vunit)
is close to that of the \fion{Al}{ix} line ($\simeq-2700$\vunit;
see Fig.~\ref{coronal2}, left panel), suggesting that the blue 
wing of the \pion{He}{i} 1.0833\mic\ line arises in the same jets
as the material responsible for \fion{Al}{ix}. If this is the case 
the deprojected velocities approach 25000\vunit. 

\begin{table*}
\caption{Swift photometry. Data are in magnitudes. 
\label{uvot}}
 \begin{tabular}{ccccccc}\hline
 Swift filter:
  &  $v$    & $b$    & $u$    & $uvw1$ &  $uvm2$ &  $uvw2$ \\
Effective $\lambda$ (\AA) ~~= &      5410 & 4321 & 3442 & 2486 & 2221  &  1991 \\
Day$^*$ & &&&&& \\ \hline
4.93 & $11.46\pm0.03$ & --- & --- & --- & $11.02\pm0.03$ & --- \\
9.49 & $13.04\pm0.03$ & $12.81\pm0.03$ & --- & $11.84\pm0.02$ &$12.07\pm0.03$ & $11.75 \pm0.02$ \\
9.65 & ---&---&---&---&---&$12.09\pm0.03$\\
10.49  &  $13.17\pm0.03$&$13.03\pm0.03$ &$12.04\pm0.03$&$12.06\pm0.02$&$12.31\pm0.03$
& $12.04\pm0.02$ \\
10.63 &---&---&---&$11.98\pm0.02$ &---&  $12.03\pm0.03$ \\
12.49 & $13.73\pm0.03$ & $13.68\pm0.03$ & $12.65\pm0.03$ & $12.53\pm0.02$ & $12.81\pm0.03$& $12.57\pm0.02$\\
12.71 &---&---&---&---&---& $12.98\pm0.03$\\
13.82 & $13.94\pm0.04
$ & $13.92\pm0.03$ & $12.89\pm0.03
$ & $12.81\pm0.03$
 & $13.03\pm0.03$
 & $12.86\pm0.02$ \\
14.03  & --- & --- & ---& --- & $12.92\pm0.03$
 & $12.67\pm0.03$ \\
23.63 &  $14.44\pm0.04
$ &  $14.63\pm0.03
$ & $13.50\pm0.03
$  &$13.40\pm0.03$
& $13.58\pm0.03
$ &  $13.43\pm0.02$ \\
27.88  &  $14.58\pm0.04
$ & $14.80\pm0.03
$ & $13.61\pm0.03
$ &  $13.57\pm0.03
$ & $13.71\pm0.03
$ & $13.59\pm0.03
$ \\
29.07  & --- &  ---    &   ---   &    --- & $13.96\pm0.04$ & $13.85\pm0.04$ \\
39.21  & $16.25\pm0.10$ & $16.21\pm0.05$ & $15.23\pm0.05$ & $15.20\pm0.05$ & $15.40\pm0.06$& $15.25\pm0.04$ \\
43.13  & $16.86\pm0.14$ & $17.15\pm0.09$ &  $16.26\pm0.07$ & $16.14\pm0.07$ & 
 $16.62\pm0.09$ & $16.19\pm0.06$ \\
46.24  &---& ---& $16.24\pm0.11$ &  $16.39\pm0.08$ & --- & --- \\
 \hline
 \multicolumn{7}{l}{$^*$Day numbers, with $t=0$ at MJD 59736.72, 
 are means of individual values.}
 \end{tabular}
\end{table*}

We attribute the feature at 2.116\mic\ to a blend of
\pion{He}{i} $^1$S--$^1$P$^o$ $\lambda=2.1138$\mic\ 
and $^3$P$^o$--$^3$S $\lambda=2.1120$\mic\ (see Fig.~\ref{data}). 
This feature is commonly seen in erupting novae 
\citep[][and references therein]{raj15}, and was present in 
the 2019 eruption of the RN V3890~Sgr \citep{evans22}. 
In the high resolution spectra, the feature is asymmetric
at the earliest epoch, with a weak blue wing superposed on a 
stronger core. Although the line weakened considerably
after $\sim$~day~12, making the properties of the two components 
difficult to measure,
fitting two gaussians to it reveals a picture that is broadly
similar to the \pion{He}{i} 1.0833\mic\ line (see Fig.~\ref{evol}), 
supporting the view that it is due to \pion{He}{i} 
\citep[rather than \fion{Ca}{ix}; see][]{evans22}.
However, there is also a distinct possibility that this feature 
has a contribution from an unidentified coronal line at $\sim2.1$\mic,
seen earlier in several classical novae, for example V1974 Cyg 
\citep{wagner96}, V959 Mon \citep*{banerjee12b}, and V5668 Sgr \citep{gehrz18}.

The presence of a \pion{He}{i} line in the 
coronal gas at temperature $7.1\times10^5$~K
is very surprising. At this temperature,
essentially all the helium is doubly-ionised \citep[see][]{arnaud85}.
This suggests that there is a region of the circumstellar
environment in which the temperature is considerably lower.
The data in \cite{arnaud85} suggest that, below $2.2\times10^4$~K,
helium is predominantly neutral, whereas between $4\times10^4$~K and 
$\sim6.3\times10^4$~K it is primarily in the form of \pion{He}{ii}.
\pion{He}{iii} predominates above $\sim8\times10^4$~K.
We conclude that, in addition to the hot ($7.1\times10^5$~K) ``coronal'' 
region, there must exist a cooler ($\ltsimeq2\times10^4$~K) region where 
the helium is neutral. We suggest that this material resides in the
torus containing the 750\vunit\ gas, postulated above.

There may be some supporting evidence for the existence of
the orbital plane torus in the Swift X-ray data, 
which imply a hydrogen absorbing column of
$N_{\rm H} \simeq3\times10^{21}$~cm$^{-2}$ along the line of sight to 
\us. Using the conversion given by \cite {liszt14} for the intersellar medium,
\[ N(\mbox{\pion{H}{i}})/E(B-V) = 8.3 \times10^{21} \mbox{cm$^{-2}$mag$^{-1}$} \:\:,\]
together with the assumed $E(B-V)=0.2$, suggests that only 
$\sim1.7\times10^{21}$~cm$^{-2}$ of this is interstellar. The 
difference between these values, $\sim1.3\times10^{21}$~cm$^{-2}$,
presumably is due to \pion{H}{i} in the circumstellar
environment. Even allowing for the uncertainty in $E(B-V)$, the difference
is at least $\sim5\times10^{20}$~cm$^{-2}$.

To  estimate the \pion{H}{i} column density in the circumstellar gas,
we assume wind density $\propto{r}^{-2}$, ejected over a period $\delta{t}$ 
\citep[$\simeq20$~days;][]{hachisu00}. The column density at time $t$ is
\[ N \simeq \frac{\dot{M}}{4\pi\overline{m}V^2} \: \:\:
\frac{\delta{t}}{t(t-\delta{t})} \:\:,\]
\citep[cf.,][]{evans22} where $\dot{M}$ is the mass ejection rate in the 
eruption and $\overline{m}$ is the mean atomic mass.
For a mass-loss rate $\sim10^{-7}$\Msun~yr$^{-1}$ \citep{hachisu00}, 
$V=750$\vunit\ for the core component (see above), and 
$\overline{m}\simeq1$,
\[ N \sim 1.2\times10^{22} \:\: \frac{\delta{t}}{t(t-\delta{t})} \mbox{~cm$^{-2}$}\:\:,\]
where both $t$ and $\delta{t}$ are in days, so that $N\sim8\times10^{20}$~cm$^{-2}$
after $\sim30$~days. Given the crudeness of the analysis, and the uncertainty
in converting $E(B-V)$ to $N_{\rm H}$, this is surprisingly close to
the estimated circumstellar contribution to $N_{\rm H}$.

\begin{figure}
 \includegraphics[width=8cm,keepaspectratio]{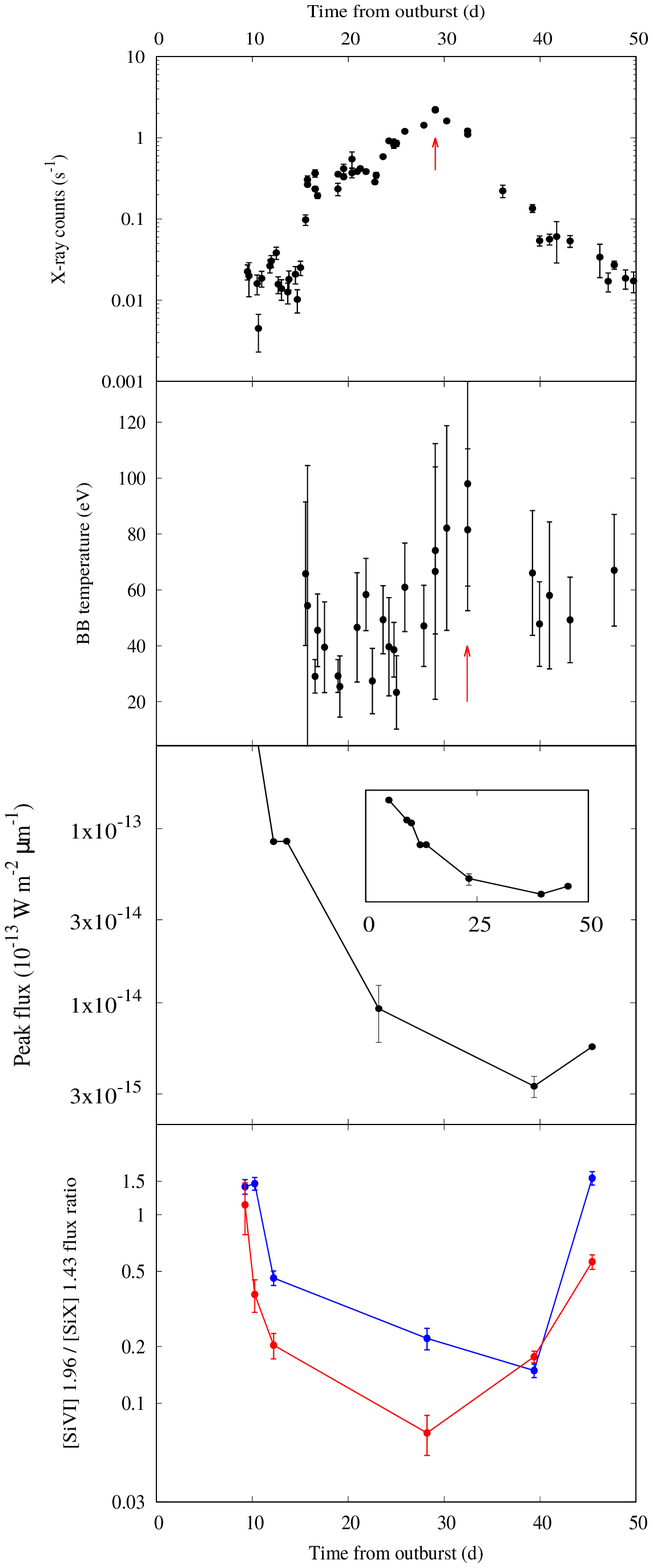}
 \caption{Top: Swift X-ray counts; arrow indicates maximum count rate.
 Upper middle: Black body temperature of the Swift source;
 arrow indicates maximum temperature.
 Lower middle: evolution of the peak flux of the 
 \pion{He}{i} 1.0833\mic\ line.  Inset shows
entire decline over the period of our observations.
  Bottom: evolution of the \fion{Si}{vi} 1.96\mic/\fion{Si}{x} 1.43\mic\
 flux ratio. Red points, fluxes of receding jets, blue points those
 of approaching jets  (see Section~\ref{coronals}).\label{min}}
\end{figure}

\subsection{The IR continuum\label{IRC}}

We have obtained $BV\!I$ photometry of \us\ from the AAVSO database. 
We also have UV/Optical Telescope (UVOT) photometry from
the Neil Gehrels Swift Observatory
\citep{gehrels04}, in the $v$ (5468\AA), $b$ (4392\AA), $u$ (3465\AA), 
$uvw1$ (2600\AA), $uvm2$ (2246\AA), $uvw2$ (1928\AA) filters; the Swift 
photometry is given in Table~\ref{uvot}.

At the shortest ($\ltsimeq1$\mic) wavelengths, the data are 
reasonably fitted by a black body function. However, for a more satisfactory fit
we require photometry at the shortest wavelength ({\it uvw}2), together 
with complementary photometry at longer wavelengths.
We select UVOT and AAVSO data within $<350$~min ($<20$\% of 
the orbital period) of the IR spectroscopy.
The datasets that satisfy these criteria are for June 16, 
June 17 and June 30 (see Fig.~\ref{SED}).
We have fitted black body curves to the UVOT/$BV\!R$ data; 
the black body temperatures, $T_{\rm BB}$, are $19000\pm2000$~K (June 16),
{$16500\pm1000$~K (June 17)}, $20000\pm1000$~K (June 30). 
We have also fitted black body curves to Swift data that do not coincide
closely with the IR data. Fig.~\ref{BB} shows the time-dependence of
$T_{\rm BB}$, its bolometric flux and luminosity, 
and the angular diameter and linear size of the emitter. 
A distance of 10~kpc has been assumed.
The temperature of the black body remained constant ($\sim20000$~K) 
over the 40~days of our IR observations, but its flux/luminosity, and 
therefore its dimensions, declined significantly. 

{\cite{evans01} suggested that the black body might be the WD itself, 
but this can not be the case, as the nova is in the ``Super Soft'' X-ray 
phase\footnote{See e.g. \cite*{page20} for an explanation.}, 
and therefore much hotter, at this time 
(see below). Neither can it be the accretion disc, the bolometric
luminosity of which in \us\ is $\sim535$\Lsun\ for an
accretion rate of \Mdot{2.5}{7} \citep{hachisu00} and a 
1.37\Msun\ WD of radius 0.01\Rsun. This luminosity is of the same order as 
that of the black body; moreover, the dimensions are comparable 
with the binary separation. However, as the disc in \us\ is seen 
almost edge-on, the observed luminosity of the disc is less than the 
accretion {luminosity}. 
Using the formalism of \cite{paczynski80} 
for a standard limb-darkened disc (i.e., one that has not been subject 
to a nova eruption), we find that the bolometric luminosity of a 
nearly-edge-on disc is fainter than that of one seen face-on by a factor 
$\sim1.1\times10^{-3}$. In any case it is very 
likely that the accretion disc is destroyed 
early in the eruption \citep{drake10}.}

{Might the black body be the secondary star? The K2IV secondary has 
radius 2.1\Rsun\ \citep{thoroughgood01};
{assuming effective temperature
$T_{\rm eff}\simeq4600$~K, typical of field K2 stars \citep{zombeck}, 
the bolometric luminosity $L_{\rm bol}\sim1.8\Lsun$}. 
Thus the black body can not be the
pristine secondary. However, in the absence of the accretion disc
(which is destroyed in the eruption),
the secondary must be irradiated by the hot WD. We estimate this 
effect using the formalism of \cite*{kovetz88}.
For a WD temperature given by $kT_{\rm WD}\simeq60$~eV 
(see Fig.~\ref{min} below), the temperature of the irradiated
secondary is $T_{\rm irr}\simeq19400$~K, with a corresponding 
luminosity $\sim545$\Lsun. The observed black body temperature is 
therefore close to that expected for an irradiated secondary, while 
the luminosity is of the same order. Given the crude nature of our 
estimates, the irradiated secondary seems the least unlikely explanation.
}

Notwithstanding the nature of the black body seen in the NIR, 
the comparison of the Swift and NIR data shows that there is
clear excess radiation in the NIR, as noted by \cite{evans01}. We interpret 
this as free-free/free-bound emission. We have fitted the excess 
with a free-free/free-bound continuum, using the observed 
H$\,\beta$ flux of $\simeq1.31\times 10^{-11}$~ergs~s$^{-1}$ cm$^{-2}$  
\citep{woodward22}, which provides an estimate of the 
continuum level. For simplicitly, we assume that the electrons 
arise from H and He only, and gas temperatures of $7.1\times10^5$~K 
(corresponding to the coronal gas, in which the He is completely
ionised) and $2.5\times10^4$~K (in which the He is
neutral). The calculation of the free-free/free-bound emission
is done separately for the two temperatures. These are included in 
Fig.~\ref{SED}. There is essentially no difference between the hotter
and the cooler free-free/free-bound emission for June 16 and 30. 
However the hotter continuum overshoots the UVOT data for
June 17; the continuum coming from the cooler gas
is in close agreement.
We tentatively suggest that the free-free continuum arises in 
the cooler region from which the \pion{He}{i} lines originate 
(see Section~\ref{hei}).
 
\subsection{Coronal line excitation}

The spectra in Fig.~\ref{data} reveal the remarkable fall and
rise of the \pion{He}{i} 1.0833\mic\ line. The dependence of its 
peak flux (continuum-subtracted) is shown in the lower middle
panel of Fig.~\ref{min} (estimating the total line flux is complicated due to
variations in the line width). 
The variation in the He  line are replicated in
the evolution of the \fion{Si}{vi} emission line at 1.965\mic,
which is seemingly absent after day 23.36, only to reappear on
day 39.55. This is demonstrated by the flux ratio 
\fion{Si}{vi} 1.96\mic/\fion{Si}{x} 1.43\mic, using
the fluxes used to determine the coronal gas temperatures 
given in Table~\ref{cor-temp}. This ratio, which gives a 
measure of the relative excitation of the gas, also reaches a
minimum at $\sim30$~days. These relationships are also shown in 
Fig.~\ref{min}, in which the upper two panels are the Swift 
X-ray counts and black body temperature of the X-ray source.
Given the relatively low cadence of our IR observations,
the data are consistent with the \pion{He}{i} line flux,
and the \fion{Si}{vi}/\fion{Si}{ix} flux ratio, both reaching
minima between days 29.38 and 39.55.

The X-ray counts reach a maximum on day 29.5. The black body 
temperature ($T_{\rm WD}$; note the distinction between this
black body temperature and that discussed in Section~\ref{IRC}) 
of the Swift source, which we identify with that of the WD,
peaks at $kT_{\rm WD}\simeq98$~eV ($T_{\rm WD}\simeq1.1\times10^6$~K), 
on day 32.9. The X-ray counts
and $T_{\rm WD}$ peak at around the same time. Thus there seems
to be a correlation between the time of the minima 
in the fluxes of the \pion{He}{i} and \fion{Si}{vi} lines,
and of the peaks in the X-ray counts and $T_{\rm WD}$. 

The behaviour of the flux ratio of the two Si lines 
(Fig.~\ref{min}; \fion{Si}{vi}, $\mbox{IP}=167$~eV;
\fion{Si}{x}), $\mbox{IP}=351$~eV) clearly demonstrates 
that the degree of ionisation peaked around
day $\sim30$, despite the fact that the black body
temperature of the Swift X-ray source rarely exceeds
$kT_{\rm WD}\sim60$~eV.

Included in Table~\ref{crit} are the ionisation
potentials of the lower ionisation stages (e.g., the ionisation 
potential of \pion{Si}{v} for \fion{Si}{vi}), and the temperatures
$T_*$ at which half the photons emitted by the corresponding  
black body can ionise the lower ion \citep[see][]{woodward22}.
The dependence of $T_{\rm WD}$ on time shows no 
value on day 5.38 (see Fig.~\ref{min}), when the X-ray counts 
were too low to provide a solution. 
However, at the earliest times, $kT_{\rm WD}\sim50$~eV, which
seems too low to photoionise and 
produce any of the high-ionisation coronal lines seen 
(e.g., \fion{Al}{ix}, \fion{Si}{vii}, \fion{Si}{x}). 

We tentatively conclude that shock ionisation and excitation 
must be operating throughout the eruption, but that photoionisation 
also contributes, possibly significantly, at around the soft 
X-ray peak when the WD temperature was at its maximum.  

In the absence of a secondary wind (as is present in the case of 
RNe with giant secondaries), presumably internal shocks 
arise in collisions between parcels  of ejecta moving at 
different velocities. However a
definitive conclusion must await a robust analysis of
the dynamics of the eruption.

\section{Conclusion}

We have presented near-infrared spectra of the RN \us, covering the period
5.4 to 45.6 days after its 2022 eruption. This is the {most
intensive} infrared observations of this object. We find that
\begin{enumerate}
 \item coronal lines were present at 9.41~days after the outburst,
 {one of}
 the earliest appearances of coronal lines in a nova, classical or recurrent;
 \item the temperature of the coronal gas was $7.1\times10^5$~K, and there
 is evidence for the simultaneous presence of much cooler 
 ($\ltsimeq2\times10^4$~K) gas;
 \item {there is evidence in the IR data for a 20000~K black body,
 which we tentatively identify with the secondary, irradiated by the hot white dwarf;}
 \item the ejecta are distributed partly in polar ``jets'' in which the coronal
 lines largely arise, and an equatorial ``torus'' in the binary plane, from
 which line emission from the cooler gas originates;
 \item for orbital inclination $82\fdg$7, jet ejection velocities as high
 as $2\times10^4$\vunit\ are implied. These are comparable with velocities
 seen in some supernovae.
\end{enumerate}

These properties mark \us\ as a truly remarkable system.

\section*{Acknowledgements}

{We thank the referee for their careful and 
thorough review of the paper.}

The Gemini observations were made possible by awards of
Director's Discretionary Time for programmes
GS-2022A-DD-105 and GN-2022A-DD-105.

The international Gemini Observatory is a program of 
NSF's NOIRLab, which is managed by the Association of Universities 
for Research in Astronomy (AURA) under a cooperative agreement 
with the National Science Foundation, on behalf of the Gemini 
Observatory partnership: the National Science Foundation 
(United States), National Research Council (Canada), 
Agencia Nacional de Investigaci\'{o}n y Desarrollo (Chile), 
Ministerio de Ciencia, Tecnolog\'{i}a e Innovaci\'{o}n (Argentina), 
Minist\'{e}rio da Ci\^{e}ncia, Tecnologia, 
Inova\c{c}\~{o}es e Comunica\c{c}\~{o}es (Brazil), 
and Korea Astronomy and Space Science Institute (Republic of Korea).

The IRTF observations were conducted under programme 2022A003.
The Infrared Telescope Facility is operated by the
University of Hawaii under contract 80HGTR19D0030 with the 
National Aeronautics and Space Administration.

CEW acknowledges partial support from NASA grant 80NSSC19K0868.
DPKB is supported by a CSIR Emeritus Scientist grant-in-aid and 
is being hosted by the Physical Research Laboratory, Ahmedabad. 
RDG was supported by the United States Airforce.
KLP acknowledges funding from the UK Space Agency.
SS acknowledges partial support from an ASU Regents' Professorship.

We acknowledge with thanks the variable star observations from the 
AAVSO International Database contributed by observers worldwide and 
used in this research.

\section*{Data availability}
The raw infrared data in this paper are available from the Gemini Observatory
Archive, https://archive.gemini.edu/, and from the IRTF archive,
http://irtfweb.ifa.hawaii.edu/research/irtf\_data\_archive.php
The Swift data are available from
https://www.swift.ac.uk/swift\_live/ and \\ 
https://heasarc.gsfc.nasa.gov/cgi-bin/W3Browse/w3browse.pl






\bsp	
\label{lastpage}
\end{document}